\documentclass{osa-article}

\journal{osac}

\articletype{Research Article}

\begin{document}

\title{Time-domain modeling of interband transitions in plasmonic systems}

\author{Max Pfeifer,  			 \authormark{1} 
        Dan-Nha Huynh,			 \authormark{1}
				Gino Wegner,  			 \authormark{1}
				Francesco Intravaia, \authormark{1}
				Ulf Peschel,  			 \authormark{2}
        and
				Kurt Busch,   			 \authormark{1,3,*}}

\address{\authormark{1}Humboldt-Universit\"{a}t zu Berlin, Institut f\"{u}r Physik, 
                       AG Theoretische Optik \& Photonik, D-12489 Berlin,Germany\\
         \authormark{2}Physikalisch-Astronomische Fakult\"{a}t, Institut f\"{u}r Festk\"{o}rpertheorie und -optik,
				               Friedrich-Schiller-Universit\"{a}t Jena, D-07743 Jena, Germany\\
         \authormark{3}Max-Born-Institut, D-12489 Berlin, Germany}

\email{\authormark{*}kurt.busch@physik.hu-berlin.de} 



\begin{abstract}
Efficient modeling of dispersive materials via time-domain simulations of the Maxwell 
equations relies on the technique of auxiliary differential equations. In this approach, 
a material's frequency-dependent permittivity is represented via a sum of rational 
functions, e.g. Lorentz-poles, and the associated free parameters are determined by 
fitting to experimental data.
In the present work, we present a modified approach for plasmonic materials that requires
considerably fewer fit parameters than traditional approaches. 
Specifically, we consider the underlying microscopic theory and, in the frequency domain, 
separate the hydrodynamic
contributions of the quasi-free electrons in partially filled bands from the interband 
transitions. As an illustration, we apply our approach to Gold and demonstrate how to
treat the interband transitions within the effective model via connecting to
the underlying electronic bandstructure, thereby assigning physical meaning to the 
remaining fit parameters. Finally, we show how to utilize this approach within the 
technique of auxiliary differential equations.
Our approach can be extended to other plasmonic materials and leads to efficient 
time-domain simulations of plasmonic structures for frequency ranges where interband 
transitions have to be considered.
\end{abstract}

\section{Introduction}
\label{intro}

In macroscopic electrodynamics at near-IR or optical frequencies, linear material 
properties are commonly described by frequency-dependent permittivities $\epsilon(\omega)$. 
Their experimental determination happens, e.g., by means of ellipsometry
\cite{Fujiwara2007}
and the so-obtained data can directly be fed into computational approaches that
solve Maxwell's equations in the frequency domain. Their incorporation into 
time-domain Maxwell-solvers requires more effort because, in the time domain, 
the constitutive relation $\vec{D}(\omega) = \epsilon_0 \epsilon (\omega) \vec{E}(\omega)$ 
translates to a convolution integral, thus turning Maxwell's equations into a 
set of numerically demanding integro-differential equations. Here, $\epsilon_0$
denotes the vacuum permittivity.\\
Within the framework of auxiliary differential equations (ADEs), this convolution 
integral is circumvented and replaced by a set of differential equations for 
auxiliary polarizations or auxiliary currents. This is accomplished 
by writing the permittivity as a sum of rational functions in $\omega$ that 
respect appropriate Kramers-Kronig relations in order to ensure causality 
\cite{Jackson1998}
(for concrete examples of ADEs, see section \ref{ADEs} below).
For instance, metals are often treated via the so-called Drude-Lorentz model
\begin{align}
  \label{drude-lorentz}
	\epsilon(\omega) = 1 - \frac{\omega_p^2}{\omega(\omega+i\gamma_{\rm D})}
	                     + \sum_{m=1}^N \frac{f_m \omega_p^2}{\omega_m^2-\omega^2 - i \omega \gamma_m}.
\end{align}
Here, the 1st and 2nd term on the r.h.s. represent, respectively, the vacuum contribution and
the (collective) response of the metal's free electrons, the so-called Drude model with plasma 
frequency $\omega_p$ and phenomenological damping constant $\gamma_{\rm D}$. \\
In dielectrics, the 3rd term on the r.h.s. of Eq.~\eqref{drude-lorentz} models electrons 
that are elastically bound to their parent nuclei so that the $f_m$ and $\omega_m$ 
$(m=1,\dots, N)$ may be interpreted, respectively, as oscillator strengths and 
frequencies of corresponding chemical bonds or may be associated with certain 
transitions in the electronic bandstructure of semiconductors. This particular 
form of the linear response has been suggested on empirical grounds by Sellmeier 
in 1871
\cite{Sellmeier:1871},
albeit for different reasons. 
In addition, a set of phenomenological damping constants $\gamma_m$ $(m=1,\dots, N)$ 
are introduced. As a result, each Lorentz pole features three adjustable parameters 
that are obtained from fits to experimental permittivity data, where the oscillator 
strengths obey a sum rule. Often, additional frequency dependencies are associated 
to the damping constants 
\cite{Djurisic:2000}.\\
In metals, Lorentz poles are used to account for interband transitions so that 
$f_m$, $\omega_m$, and $\gamma_m$ ($m=1,\dots,N$) and the number $N$ of Lorentz 
poles represent effective parameters without straightforward physical interpretation. 
Nonetheless, sophisticated and rather cumbersome methods
\cite{Gharbi:2020}
have been developed to fit the parameters of the Drude-Lorentz model, Eq.~\eqref{drude-lorentz},
to experimental data such as those provided by the classic measurements of 
Johnson \& Christy for noble metals
\cite{Johnson:1972}
or the data reported in the handbook of optical constants of solids edited by Palik
\cite{Palik1998}.\\
In the present work, we adopt a more microscopic viewpoint on $\epsilon (\omega)$ 
and demonstrate (i) how this leads to a considerable reduction in the number of fit 
parameters and (ii) how the remaining parameters may be interpreted physically.
Further, our approach provides an alternative perspective on Eq.~\eqref{drude-lorentz} 
and offers a systematic way of improving the accuracy of time-domain computations 
without introducing additional fit parameters. Our work is organized as follows.
In section \ref{theory-metals}, we review some of the available literature, develop 
our approach, and apply it to the permittivity of gold, the perhaps most frequently 
used plasmonic material.
We proceed in section \ref{time-domain} to the adaptation of our approach for
time-domain simulations and provide numerical assessments of its validity. 
In section \ref{conclusions}, we summarize our findings.

\section{Theory of the frequency-dependent permittivity of metals}
\label{theory-metals}

One of the earliest discussion on the permittivity of metals dates back to two works 
of Darwin
\cite{Darwin:1934,Darwin:1943}.
There, on a classical level, it was explained that the bound and free electron response 
are additive and that there is no local field correction in a plasma.
On the quantum mechanical level, Bohm and Pines
\cite{Bohm:1951,Pines:1952,Bohm:1953,Pines:1953}
have elaborated the physical properties of metals. Regarding the metals' response to 
electromagnetic fields
\cite{Bohm:1953},
they have demonstrated that the complicated multi-particle interactions in a metal can
be classified into two types which require rather distinct treatments. More precisely,
there exist collective excitations of the electron system that are dominant at long
wavelengths and that can be
treated via Fermi-liquid theory (Landau-Silin theory
\cite{Nozieres1999}) 
where the periodic (i.e. atomic) nature of the ionic background is essentially irrelevant. 
Through a corresponding canonical transformation of the original Hamiltonian, the 
remaining short-range interactions are then mapped onto a collection of independent 
particles that move in an effective periodic potential. Consequently, this separation 
of spatial scales leads to a description of the electron system as a mixture of a 
Fermi liquid and a set of non-interacting effective particles.\\
It is to this mixture that the aforementioned arguments of Darwin may be applied so
that the linear permittivity of metals may be decomposed into
\begin{align}
	\label{fermi-plus-particles}
	\epsilon_{\rm Metal} & = 1 + \chi_{\rm FL} (\omega) + \chi_{\rm EP} (\omega), 
\end{align}
where, $\chi_{\rm FL} (\omega)$ and $\chi_{\rm EP} (\omega)$ denote the linear 
susceptibilities of the Fermi liquid and the effective particles, respectively.

\subsection{Drude model and extensions}
\label{drude-model}

The Fermi liquid contributions can be treated in numerous ways with varying degrees
of sophistication. The aforementioned Drude model 
\begin{align}
	\label{drude}
	\epsilon_{\rm D}(\omega) = 1 + \chi_{\rm D} (\omega)
	                         = 1 - \frac{\omega_p^2}{\omega(\omega-i\gamma_{\rm D})}
\end{align}
represents the simplest approach that works rather well for simple metals and can 
even be derived from purely classical arguments
\cite{Darwin:1934}
where the plasma frequency $\omega_p^2 = n e^2/(\epsilon_0 m)$ is expressed in terms
of the density $n$, the charge $e$, and mass $m$ of electrons. In some cases the `1'
in Eq.~\eqref{drude} is replaced by $\epsilon_\infty$ to account for
a background polarization.
In an actual Fermi liquid $n$ and $m$ need to be replaced by density and effective 
mass of the quasi-particles. 
Since, however, $\omega_p$ is usually determined by fitting $\epsilon_{\rm D} (\omega)$ 
to experimental data (see 
\cite{Olmon:2012}
for a recent example of fitting the Drude permittivity of gold),
in practice, this difference does not matter too much.\\
For strongly correlated systems such as the ferromagnetic metals nickel, cobalt 
or iron, the Drude model must be extended to better account for the electronic 
correlations
\cite{Wolff:2013}.\\
Further, if other relevant length scales such as the size of a metallic particle 
or the distance between two metallic particles become comparable to the electrons' 
mean free path, nonlocal effects such as Landau damping (which already occurs in 
classical theories) and 
the Fermi pressure (a purely quantum mechanical effect) as well as the detailed behavior 
of the liquid near surfaces have to be taken into account
\cite{Mortensen:2021}. Since the Fermi liquid 
contributions are not at the focus of our work, we would like to refer to the recent 
literature
\cite{Raza:2015,Toscano:2015,Ciraci:2016,Moeferdt:2018,Reiche:2020} and references therein
for details.

\subsection{Interband transitions}
\label{interband}

The effective particle susceptibility can be obtained from standard linear response
theory. To do so, it is important to note that the effective particle moves in a
periodic potential which leads to the formation of an effective energy bandstructure
that is related to the actual electronic bandstructure in the following way. First,
for a metal the conduction band is partially filled and the electrons that occupy
these states represent the free electrons that are treated via Fermi liquid theory
as described in the previous section \ref{drude-model}. 
Consequently, as the aforementioned canonical transformations `transform away' the
Fermi liquid, i.e. the intraband contributions, the effective particle corresponds 
to transitions from bands that are completely filled to bands that are at least partially
filled. In other words, to the level of approximation discussed above, the 
effective particle contributions to the susceptibility are associated with interband
transitions of an (effective) and potentially highly doped semiconductor. 
Wysin et al.
\cite{Wysin:2013}
have adopted this viewpoint in order to determine the contribution of these interband
transitions on the Faraday rotation in metallic nanostructures. The corresponding 
contribution of the hydrodynamic part has been developed in Ref.
\cite{Wolff:2013}.\\
Following the above prescription, the contributions of the effective particles to the
susceptibility can be determined via linear response theory under dipole coupling. This
gives
\begin{align}
	\label{chi}
	\chi_{\rm EP} (\omega) = \frac{2 n}{\hbar}
	                         \sum_{l \vec{k}} \sum_{l^\prime \vec{k}^\prime} 
													 \frac{p_{l \vec{k}} \, \omega_{l l^\prime \vec{k} \vec{k}^\prime}
												         \vert 
															      \langle l \vec{k} \vert \hat{\vec{d}} \vert l^\prime \vec{k}^\prime \rangle
											           \vert^2}{
															   \omega_{l l^\prime \vec{k} \vec{k}^\prime}^2
														     + \gamma_{l l^\prime \vec{k} \vec{k}^\prime}^2
																 + 2i \gamma_{l l^\prime \vec{k} \vec{k}^\prime} \omega
																 - \omega^2
															   }.
\end{align}
Here, $p_{l \vec{k}}$ and 
$\omega_{l l^\prime \vec{k} \vec{k}^\prime} = (\epsilon_{l \vec{k}}-\epsilon_{l^\prime \vec{k}^\prime})/\hbar$
denote, respectively, the occupation of a Bloch state (labeled by band index $l$ and 
wave vector $\vec{k}$) of the effective particle with energy $\epsilon_{l \vec{k}}$ and 
the transition frequency between two Bloch states. In addition, we have introduced 
phenomenological damping constants $\gamma_{l l^\prime \vec{k} \vec{k}^\prime}$ for 
the respective transitions. 
Finally, $\langle l \vec{k} \vert \hat{\vec{d}} \vert l^\prime \vec{k}^\prime \rangle$ is the
matrix element of the electric dipole operator $\hat{\vec{d}}$ between two Bloch states. For
the details of the calculations leading to \eqref{chi}, we refer to the work of Wysin et al.
\cite{Wysin:2013}.
%
%

\subsection{Parabolic two-band model for Gold}
\label{2bands}

\begin{figure}[h!]
	\centering\includegraphics[width=9cm]{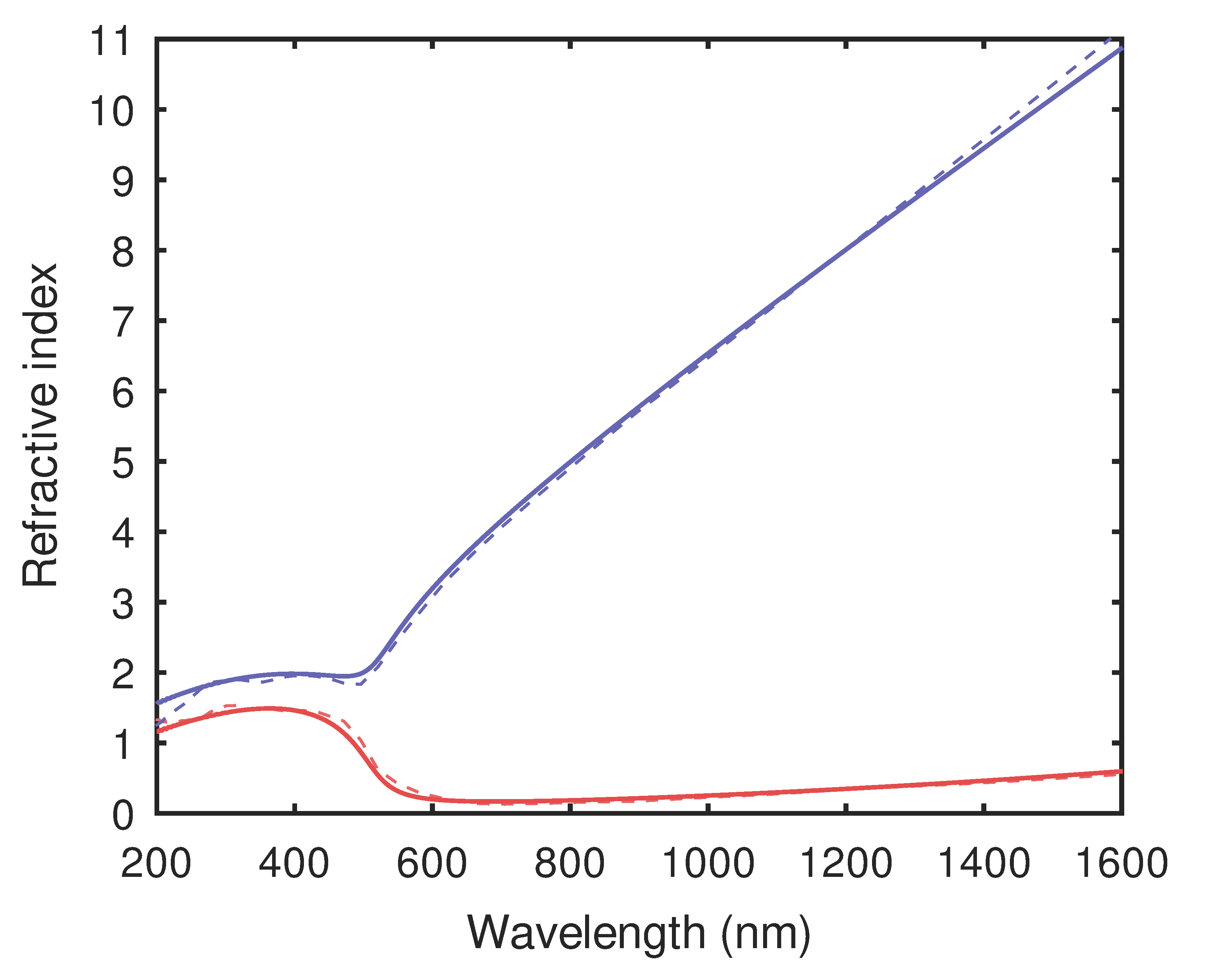}
	\caption{\label{johnson-fit}
	         Johnson \& Christy data (dashed lines) for the complex refractive index 
					 (real part: blue lines, imaginary part: red lines) of Gold in the range between 
					 1600nm and 200nm as well as best fit (solid lines) of the parabolic 
					 two-band model via Eqs.~\eqref{fermi-plus-particles}, 
					 \eqref{drude}, and \eqref{chi1D}. 
					 The fit has been done for the range of 1937nm to 187nm. This is approximately
					 the range considered in ref. 
					 \cite{Gharbi:2020}. 
					}
\end{figure}

We now apply the framework outlined in Eqs.~\eqref{fermi-plus-particles}, \eqref{drude}, 
and section \ref{interband} to an actual metal -- we chose the arguably most important metal
for applications in plasmonics: gold. To do so, it is important to recall the results 
of corresponding electronic bandstructure computations. 
Specifically, electronic bandstructure theory reveals that gold features a fully occupied 
5d-band that, energetically, lies below the partially filled 6sp-band
\cite{Ngoc:2015} 
where the energy difference for direct optical transitions is between 1.8 and 2.45 eV,
thus covering the frequency range from the red to the green/cyan range of the optical
spectrum.
Consequently, the electrons of the partially filled 6sp-band correspond to the Fermi
liquid discussed above. 
Symmetry considerations stipulate that the dipole matrix elements between the 5d- and
the 6-sp band are strongly suppressed. However, the 5d-band exhibits a strong van Hove 
singularity near the X-point of the Brillouin zone which compensates the small dipole 
matrix elements. 
Consequently, if we limit our considerations of the permittivity to the optical 
frequency range, we can restrict the effective particle susceptibility in Eq.~\eqref{chi} 
to a two-band model that only includes direct optical transitions, $\vec{k} = \vec{k}^\prime$. 
As the transitions are appreciable only near the X-point, and given that the 5d-band 
is much flatter in $\Gamma$-X direction than in any other direction, we may adopt 
a parabolic approximation to both bands and limit the remaining 
summation in Eq.~\eqref{chi} to an integration along the relevant part of the 
$\Gamma$-X line. For this effective two-band semiconductor, the effective Fermi 
energy represents a fitting parameter that lies between the top of the lower and 
the bottom of the upper band. 
Assuming a step function for the occupation of the effective two-band model and
further assuming that the corresponding dipole matrix elements and damping constants 
do not depend on wave vector, Eq.~\eqref{chi} evaluates to (for details of the 
calculation using the same notation, see Ref. 
\cite{Wysin:2013})
\begin{align}
	\label{chi1D}
  \chi_{\rm EP} (\omega) & = Q \int_0^{s_{\rm U}} {\rm d}s \, 
	                             \frac{2s^2}{\left( \omega_g + s^2 \right)
													                 \left( \left( \omega_g + s^2 \right)^2
																			        - \left( \omega + i \gamma \right)^2   
																			     \right) } \\
	\label{Q1D}
	Q & = \frac{2ne^2 \vert M \vert^2 \hbar}{\epsilon_0 m_e^2} 
	      \cdot
		  	\frac{2V s_{\rm U}^2}{\pi^3}
		   	\cdot
		  	\left( 
		   		\frac{\hbar}{2m^*}
		  	\right)^{1/2}
\end{align}
Here, we have introduced the scaled wave vector $s = (\hbar/2m^*)k$ (with dimension
1/time$^{1/2}$), the upper limit $s_{\rm U}$ of the scaled wave vector integration,
and have denoted the dipole matrix elements with $M$. The energies are measured from the 
top of the effective valence band, so that the dispersion of the parabolic bands are 
$\epsilon_{l \vec{k}} = - \hbar^2 k^2/2m_v $ and 
$\epsilon_{l^\prime k^\prime} = \hbar \omega_g + \hbar^2 k^2/2 m_d$, where $m_h$ and
$m_e$ represent the curvatures of the effective valence and the effective conduction band
around the Fermi surface, 
respectively. The reduced mass is defined via $1/m^* = 1/m_h + 1/m_e$. The above integral 
can be evaluated exactly in terms of a sum of arc tangents. However, as we shall discuss 
in the next section, this does neither provide further physical insight nor does it 
help with formulating ADEs for time-domain simulations. \\
Combining Eqs.~\eqref{fermi-plus-particles}, \eqref{drude}, and \eqref{chi1D}, we now a
have a model for the linear permittivity of gold with a total of 6 fitting parameters:
$\omega_p$, $\gamma_{\rm D}$, $Q$, $\omega_g$, $\gamma$, and the integration limit
$s_{\rm U}$.
We expect that this model covers the low frequencies all the way to the blue end of the
visible frequency range. In addition, we expect that the gap frequency $\omega_g$ of the
effective two-band model is close to the aforementioned energetic separation between
the 5d- and the 6-sp band.\\
Upon carrying out the above-sketched fitting procedure to the experimental data for
gold by Johnson \& Christy
\cite{Johnson:1972}
via a standard Monte-Carlo approach, we obtain the values 
$\omega_p = 1.32 \cdot 10^{16}$ Hz,
$\gamma_{\rm D} = 1.23 \cdot 10^{14}$ Hz,
$Q = 2.72 \cdot 10^{24} $ 1/Hz$^{3/2}$,
$\omega_g = 3.63 \cdot 10^{15}$ Hz,
$\gamma = 2.41 \cdot 10^{14}$ Hz, and
$s_{\rm U} = 2.66 \cdot 10^{8}$ Hz$^{1/2}$. 
In Fig.~\ref{johnson-fit}, we display a comparison between the experimental data and
our fit using Eqs.~\eqref{fermi-plus-particles}, \eqref{drude}, and \eqref{chi1D}.
The above gap frequency corresponds to an energy of 2.39 eV and, thus, is consistent 
with expectation. The fit shows very good agreement with the experimental data 
for wavelengths up to and including the violet part of the visible spectrum. For even 
shorter wavelengths, deviations between the experimental data and the fit become more 
pronounced. Obviously, the fit could be improved by considering transitions between, 
e.g., the 4d-band and the 6-sp band thereby extending the two-band model to a multi-band 
model.
Nontheless, we would like to emphasize that the above results have been obtained with
6 fitting parameters while corresponding fits of comparable quality using 
Eq.~\eqref{drude-lorentz} require at least 5 Lorentz terms and the usage of an
$\epsilon_\infty$ instead of `1' (see section \ref{drude-model}), totalling at least 
16 fitting parameters 
\cite{Gharbi:2020}.

\section{Time-domain modeling of interband transitions}
\label{time-domain}

As described above, in time-domain numerical schemes of Maxwell's equations 
frequency-dependent permittivities are usually treated via ADEs. Unfortunately, 
while Eq.~\eqref{drude} is amenable to an ADE approach, Eq.~\eqref{chi1D} is 
not. And neither is its closed-form solution in terms of a sum of arc tangents. 
Therefore, in this section, we will develop an appropriate scheme for handling 
Eq.~\eqref{chi1D} via ADEs to any desired accuracy without introducing additional 
fitting parameters.

\subsection{Discretization of the parabolic two-band model}
\label{ADEs}

The specific form of Eq.~\eqref{chi1D} suggests that we discretize the integral
via a Gauss quadrature rule according to
\begin{align}
	\label{gauss}
  \int_a^b {\rm d}x \, f(x) & \approx \frac{b-a}{2} \, \sum_{i=1}^{N_{\rm G}}
	                                                      w_i \, f(u_i).
\end{align}
Here, $u_i = (a+b)/2 + x_i(b-a)/2$ with $i=1,\dots,N_{\rm G}$ denote the Gauss quadrature 
nodes which are expressed in terms of the roots $x_i$ of the Legendre polynomial 
$P_{N_{\rm G}}(x)$ of order $N_{\rm G}$ within the interval $[-1,1]$ 
\cite{Press1986}. 
In addition, the weights $w_i$ are given $w_i = -2/((1-x_i^2) (P_{N_{\rm G}}^\prime (x_i)^2)$
so that Eq.~\eqref{gauss} integrates exactly polynomials of degree $2N_{\rm G}-1$
\cite{Press1986}. \\
Upon applying the above Gauss quadrature to Eq.~\eqref{chi1D}, we obtain
\begin{align}
	\label{chi1D-gauss}
		\chi_{\rm EP} (\omega) & \approx \sum_{m=1}^{N_{\rm G}} \frac{a_m^2}{c_m^2 - (\omega+i\gamma)^2}, \\
	\label{gauss-parameters}
											 a_m & = \sqrt{\frac{Q \, s_{\rm U} \, s_m^2 w_m}{c_m}}, \quad 
	                       c_m = \omega_g + s_m^2, \quad 
		                     s_m = \frac{s_{\rm U}}{2}  \left( x_m +1 \right). 
\end{align}
With the above discretization \eqref{chi1D-gauss} of the single particle contribution
within the parabolic two-band model, Eq.~\eqref{chi1D}, we end up with a set of effective
Lorentz poles which can be included into time-domain simulations of Maxwell's equations 
via ADEs. The number $N_{\rm G}$ of these Lorentz poles controls the level of accuracy
with which we are approximating Eq.~\eqref{chi1D}. However, contrary to the standard
Drude-Lorentz approximations based on Eq.~\eqref{drude-lorentz} or Lorentz-pole-only
approximations 
\cite{Gharbi:2020},
our approximation scheme, Eqs.~\eqref{fermi-plus-particles}, \eqref{drude}, and
\eqref{chi1D-gauss}, does not introduce any additional fit parameters when increasing
the number of Lorentz poles (or, equivalently, when increasing the accuracy of the
approximation). 
\begin{figure}[b]
	\centering{\includegraphics[width=0.48\linewidth]{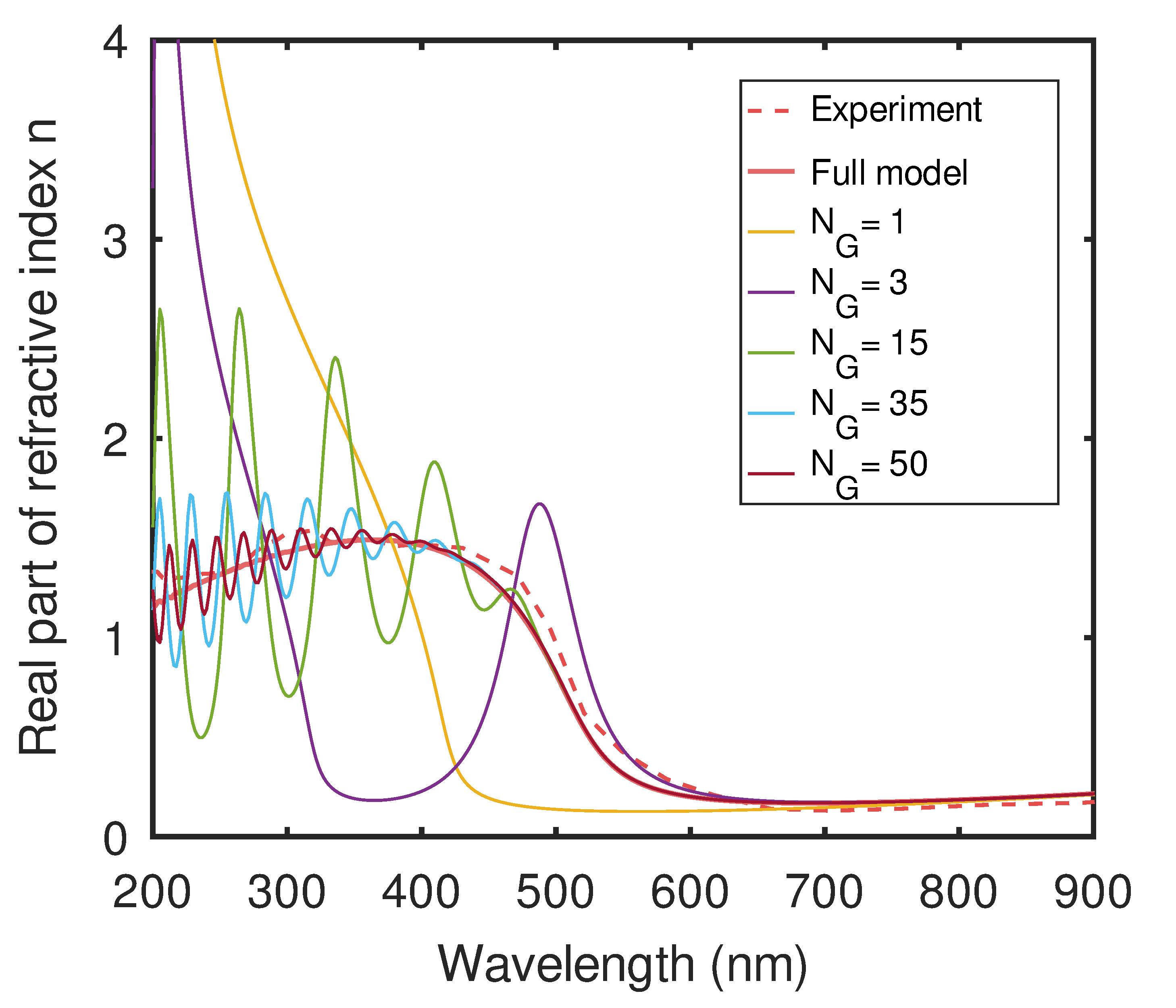}
	           \hfill
						 \includegraphics[width=0.48\linewidth]{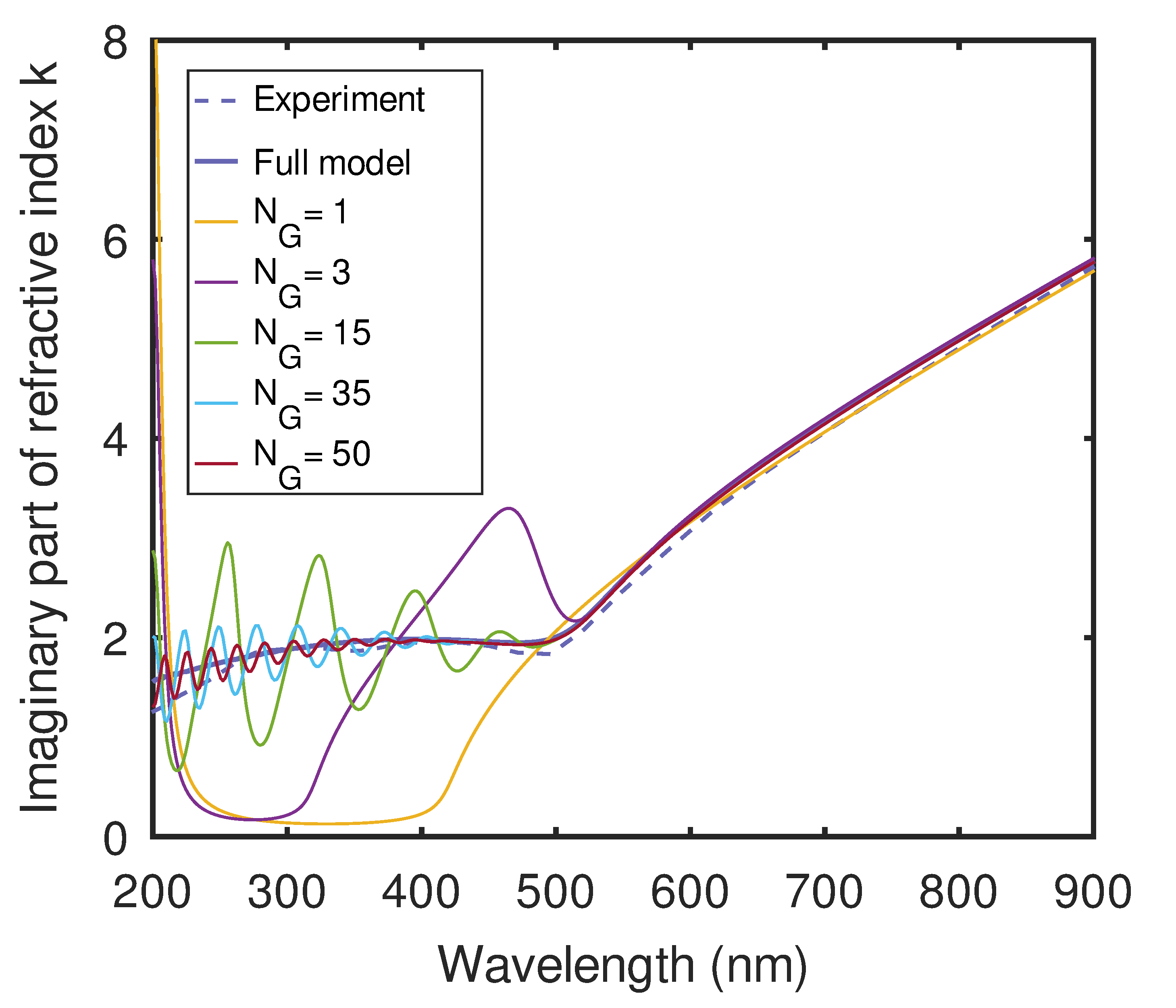}
	          }
	\caption{\label{effective-Lorentz-accuracy}
	         Complex refractive index of Gold for different Gauss quadratures
					 of the parabolic two-band model in the range between 200nm and 
					 900nm.  
					 Left panel: Real part of the refractive index.
					 Right panel: Imaginary part of the refractive index.
					 The free interband parameters $Q$, $\omega_g$, $\gamma$, and $s_{\rm U}$ have
					 been determined by fitting Eqs.~\eqref{fermi-plus-particles}, 
					 \eqref{drude}, and \eqref{chi1D} to the experimental data of
					 Johnson \& Christy (dashed lines).
					 When the number $N_{\rm G}$ of Lorentz-poles (i.e., Gauss-quadrature points) 
					 increases, both, real and imaginary part, converge to the full model. 
	        }
\end{figure}
In Fig.~\ref{effective-Lorentz-accuracy}, we depict details of the accuracy of our
effective Lorentz-pole discretization for Gold. A single effective Lorentz pole 
works very well for wavelengths larger than about 600nm, an approximation
based on three effective Lorentz poles works well down to about 550nm. For smaller 
wavelengths a considerably larger number of Lorentz poles is required for obtaining 
accurate results. In particular, by its very nature, the Lorentz pole approximation 
leads  to unphysical peaks both, in the real and imaginary part of the complex refractive
index which only become less pronounced when the number of effective Lorentz poles
is increased. Therefore, considerable care must be exercised when designing Gold-based
plasmonic elements for operation below 600nm.
Nonetheless, we would like to point out that independent of how many effective
Lorentz poles we use, we still have only the four fit parameters discussed above.
Using only these four fit parameters, we have in hand a systematic way of improving 
the accuracy of our modeling of the dielectric properties of Gold via `simply'
increasing the number of effective Lorentz poles, thus allowing us to avoid 
unsubstantiated or unphysical designs and results.
		
\subsection{Alternative Discretization of the parabolic two-band model}
\label{ADEs-alternative}
\begin{figure}[t]
	\centering{\includegraphics[width=0.48\linewidth]{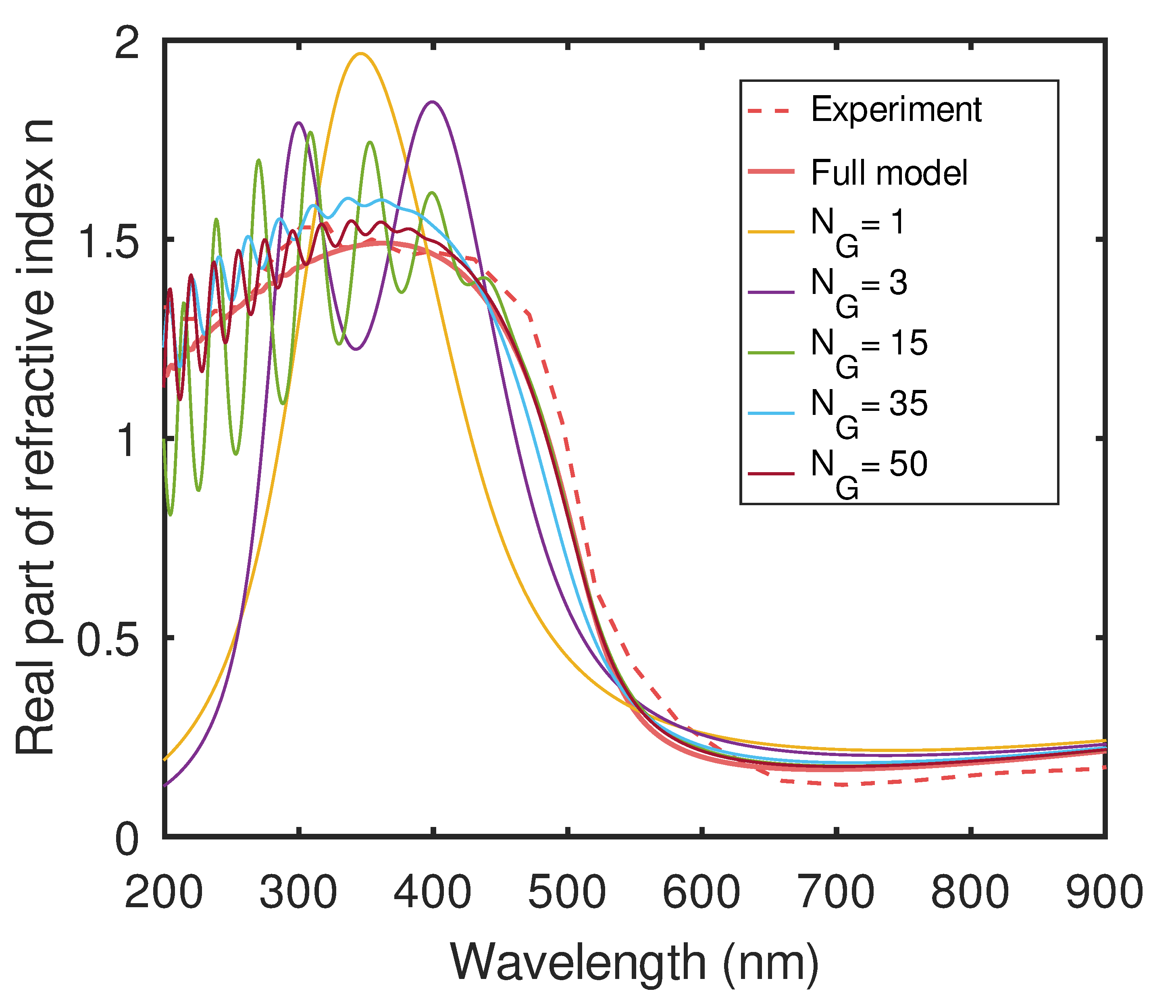}
	           \hfill
						 \includegraphics[width=0.48\linewidth]{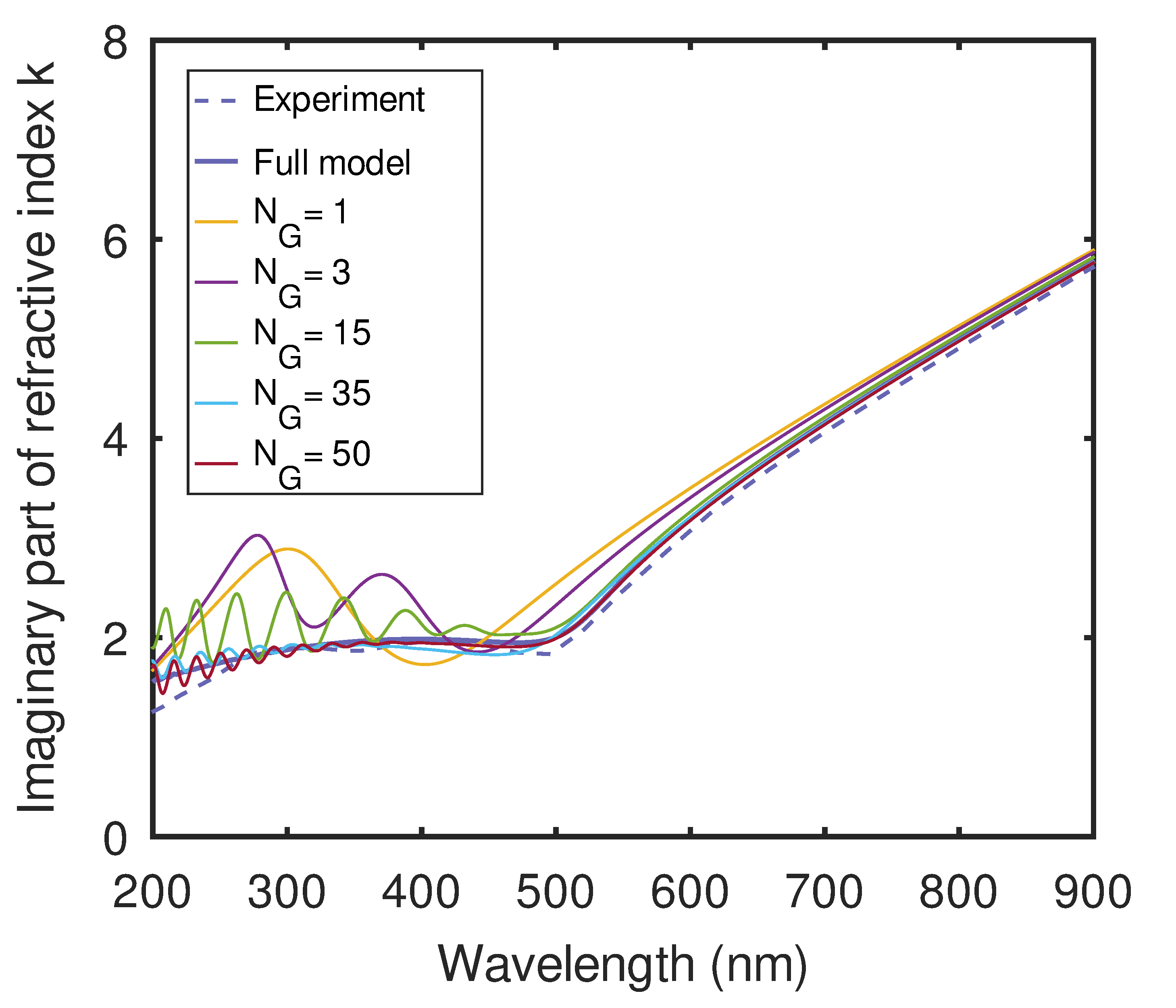}
	          }
	\caption{\label{effective-Lorentz-fit}
	         Complex refractive index of Gold for the alternative approach to Gauss 
					 quadrature approximations of the parabolic two-band model in the range 
					 between 200nm and 900nm.  
					 Left panel: Real part of the refractive index.
					 Right panel: Imaginary part of the refractive index.
					 Contrary to the results depicted in Fig.~\ref{effective-Lorentz-accuracy},
					 in this approach IBT-integral Eq.~\eqref{chi1D} is first discretized 
					 with a Gauss quadrature scheme with $N_{\rm G}$ points (i.e. Lorentz poles) 
					 according to Eq.~\eqref{chi1D-gauss}. Then, for this choice of $N_{\rm G}$,
					 the free parameters $Q$, $\omega_g$, $\gamma$, and $s_{\rm U}$ are determined
					 by fitting to the experimental data of Johnson \& Christy (dashed lines).
					 When the number $N_{\rm G}$ of Lorentz-poles (i.e., Gauss-quadrature points) 
					 increases, both, real and imaginary part, converge to the full model. 
					 An approximation with $N_{\rm G}=3$ Lorentz poles works well for wavelengths
					 larger than 450nm and a significantly larger number of Lorentz poles is
					 required for obtaining accurate results below 450nm. The values for the
					 fit parameters are listed in Tab.~\ref{fit-parameters}.
	        }
\end{figure}

Based on the above considerations, we consider a variant of the above approach. 
For a given $N_{\rm G}$, we may regard Eqs.~\eqref{fermi-plus-particles}, 
\eqref{drude}, and \eqref{chi1D-gauss} as a replacement of Eq.~\eqref{drude-lorentz} 
with fixed Drude parameters $\omega_p$ and $\gamma_{\rm D}$ and a set of four
fit parameters $Q$, $\omega_g$, $\gamma$, and $s_{\rm U}$. We then proceed
to fit the resulting approximation to the experimental data. Note, that in this
approach, for different values of $N_{\rm G}$, we obtain (slightly) different values 
for these four fit parameters. 
In Fig.~\ref{effective-Lorentz-fit} we depict the results of such a strategy for 
different numbers $N_{\rm G}$ of Lorentz poles. The corresponding values of the
fit parameters are summarized in Tab.~\ref{fit-parameters}.

\begin{table}[h!]
	\centering
	\begin{tabular}{ |c|c|c|c|c| } 
		\hline 
		$N_{\rm G}$ & $Q$ [1/Hz$^{3/2}$] & $\omega_g$ [Hz] &  $\gamma$ [Hz] & $s_{\rm U}$ [Hz$^{1/2}$]\\ 
		\hline
		1           &    1.48*10$^{27}$    & 1.00*10$^{15}$  & 5.72*10$^{15}$ &      9.52*10$^{6}$      \\ 
		\hline
		3           &    5.69*10$^{24}$    & 5.79*10$^{14}$  & 4.10*10$^{15}$ &      5.49*10$^{7}$      \\ 
		\hline
		15          &    3.16*10$^{24}$    & 2.79*10$^{14}$  & 3.67*10$^{15}$ &      8.36*10$^{7}$      \\
		\hline 
		35          &    3.16*10$^{24}$    & 3.62*10$^{14}$  & 3.77*10$^{15}$ &      1.18*10$^{8}$      \\
		\hline 
		50          &    2.83*10$^{24}$    & 2.72*10$^{14}$  & 3.65*10$^{15}$ &      1.48*10$^{8}$      \\ 
		\hline
	\end{tabular}
	\caption{\label{fit-parameters}
	         Values of the fit parameters for the alternative discretization
	         of the parabolic two-band model for Gold. These values lead to the
					 results of the complex refractive index depicted in 
					 Fig.~\ref{effective-Lorentz-fit}.
	        }
\end{table}
The results of Fig.~\ref{effective-Lorentz-fit} show that this approach leads to
a further improvement of the approximation. The single effective Lorentz pole
approximation now works well for wavelengths larger than 550nm, while the 
approximation based on three effective Lorentz poles works well down to about
450nm. For even smaller wavelengths, a significantly larger number of effective
Lorentz poles is required for accurate results where the discretization converges
to the results displayed in Fig.~\ref{johnson-fit}. However, compared with the original
discretization discussed in Sec.~\ref{ADEs}, the unphysical peaks in the complex 
refractive index are less pronounced within this alternative discretization
scheme of the IBT-integral, thus somewhat alleviating but certainly not eliminating
the afore-discussed problem of unphysical peaks in the complex refractive index. 

\section{Conclusions}
\label{conclusions}
In summary, we have developed an efficient model for the linear dielectric response
of simple metals with an emphasis on treating interband transitions and the model's
incorporation into time-domain Maxwell solver. We have illustrated this approach
by way for arguably the most important plasmonic material, Gold.\\
Our model is based on the notion of separating the metal electron's hydrodynamic 
(intraband) characteristics from the interband transitions by way of an appropriate 
canonical transformation. The resulting effective modeling of interband transitions
leads to an excellent agreement with experimental data by fitting only four free
parameters. \\
In addition, we have demonstrated two close connected approaches how this model can 
be efficiently discretized for time-domain Maxwell simulations via an arbitrary number 
of effective Lorentz poles, thereby providing a systematic way of improving the 
results without introducing additional fit parameters. With three effective Lorentz 
poles these models work well for wavelengths larger than about 450nm and significantly
more effective Lorentz poles are required for accurate results below 450nm. In this
latter wavelength range, peaks associated with the effective Lorentz poles may lead
to unphysical results. Consequently, the aforementioned systematic way of improving 
the discretization (which reduces the amplitudes of the Lorentz peaks) together with 
the fact that in all instances only six parameters -- two for the Drude permittivity 
and four parameters for the interband transitions -- represents a useful tool for 
numerical computations.
Furthermore, we should like to mention that our model naturally allows more advanced
treatments of the hydrodynamic (intraband) contributions well beyond the Drude model. 
Specifically, the Drude permittivity may simply be replaced by any of its nonlocal 
extensions
\cite{Mortensen:2021}. 
As a result, our model will be very useful for accurate modeling of
plasmonic nanostructures in regimes where both nonlocal and interband transitions 
are important.
 
\section{Backmatter}

\begin{backmatter}

\bmsection{Funding}
\noindent
German Research Foundation (DFG), Collaborative Research Center 1375 {\it Nonlinear Optics down to Atomic Scales (NOA)},
(project number 398816777).

\bmsection{Acknowledgments}
\noindent
The authors thank the Deutsche Forschungsgemeinschaft (DFG) for funding the project (project number 398816777) 
within the framework of the CRC 1375 NOA.

\bmsection{Disclosures}
\noindent 
The authors declare no conflicts of interest.

\bmsection{Data availability} 
\noindent
Data underlying the results presented in this paper are not publicly available at this time 
but may be obtained from the authors upon reasonable request.

\end{backmatter}







\end{document}